\documentclass[journal,10pt]{IEEEtran}

\usepackage{url,comment,cite}
\usepackage{bbm,dsfont}
\usepackage{amsmath,amssymb,amsfonts}
\usepackage{graphicx}
\usepackage{epsfig,epsf,psfrag,textcomp,tabularx}
\usepackage[usenames,dvipsnames]{xcolor}
\usepackage{subfigure}
\usepackage{float}
\usepackage{stfloats,mathtools}
\usepackage[algoruled,medskip,dontprintsemicolon,linesnumbered,Algorithm]{algorithm}
\usepackage{algorithmicx}
\usepackage[noend]{algpseudocode}
\usepackage{multirow}

\newcommand{\Bc}{\mathcal{B}}
\newcommand{\Gc}{\mathcal{G}}
\newcommand{\Vc}{\mathcal{V}}

\newcommand{\Kc}{\mathcal{K}}

\newcommand{\deltav}{\boldsymbol{\delta}}

\newcommand{\varthetav}{\boldsymbol{\vartheta}}

\newcommand{\varphiv}{\boldsymbol{\varphi}}
\newcommand{\psiv}{\boldsymbol{\psi}}

\newcommand{\av}{{\bf a}}

\newcommand{\dv}{{\bf d}}
\newcommand{\ev}{{\bf e}}

\newcommand{\wv}{{\bf w}}
\newcommand{\vv}{{\bf v}}
\newcommand{\xv}{{\bf x}}

\newcommand{\zv}{{\bf z}}

\newcommand{\Am}{{\bf A}}

\newcommand{\Em}{{\bf E}}

\newcommand{\Hm}{{\bf H}}
\newcommand{\Id}{{\bf I}}

\newcommand{\Ym}{{\bf Y}}

\newcommand{\EE}{\mathbb{E}} 

\def\Herm{^\mathsf{H}}
\def\Tran{^\mathsf{T}}

\newcommand{\ee}{{\rm e}}
\newcommand{\jj}{{\rm j}}  

\newcommand{\Fig}[1]{Fig.~\ref{fig:#1}}
\newcommand{\Sec}[1]{Sec.~\ref{sec:#1}}
\newcommand{\Tab}[1]{Tab.~\ref{tab:#1}}
\newcommand{\Eq}[1]{(\ref{eq:#1})}

\newcommand{\ind}[1]{\mathds{1}_{#1}}

\begin{document}

\title{Mmwave Beam Management in \\Urban Vehicular Networks
}

\author{Zana~Limani~Fazliu,~\IEEEmembership{Member,~IEEE,}
Francesco~Malandrino,~\IEEEmembership{Senior Member,~IEEE,}
Carla~Fabiana~Chiasserini,~\IEEEmembership{Fellow,~IEEE,} 
 Alessandro~Nordio,~\IEEEmembership{Member,~IEEE} %
\thanks{Z.~Limani~Fazliu is with  University of Prishtina, Prishtina,
Kosovo. C.~F.~Chiasserini is with  Politecnico di Torino  and
with CNIT, Italy. C.~F.~Chiasserini, F.~Malandrino, and A.~Nordio are with CNR-IEIIT, Italy. 

This work has been performed in the framework of the European Union's
Horizon 2020 project 5G-CARMEN co-funded by the EU under grant
agreement No 825012. The views expressed are those of the authors and
do not necessarily represent the project. The Commission is not liable
for any use that may be made of any of the information contained
therein. This work was partially supported by the Academy of Arts and
Sciences of Kosovo.

}
}

\maketitle

\begin{abstract}
  Millimeter-wave (mmwave) communication represents a potential
  solution to capacity shortage in vehicular networks. However,
  effective beam alignment between senders and receivers requires
  accurate knowledge of the  vehicles' position  for fast beam
  steering, which is often impractical to
  obtain in real time. We address this problem by leveraging
   the traffic signals regulating 
  vehicular mobility:  as an example, we may coordinate beams with red
  traffic  lights, as they correspond to higher vehicle densities and
  lower speeds. To evaluate our intuition, we propose a
 tractable, yet accurate, 
  mmwave communication model accounting for both the distance and the
  heading of vehicles being served.  Using such a model, we optimize
 the  beam design and define a low-complexity,  heuristic strategy. For increased realism, we consider
  as reference scenario a large-scale, real-world mobility trace
  of vehicles in Luxembourg. The results show that our
  approach closely matches the optimum and always outperforms static beam design based on road topology
  alone. Remarkably, it also yields better performance than 
   solutions based on real-time mobility information.
\end{abstract}

\section{Introduction}
\label{sec:intro}

High-definition maps, their real-time updates, and on-board multimedia
systems are just a few of the applications that make automated vehicles prime consumers of network
traffic. Indeed, automotive services -- both safety and 
entertainment -- are among the reference use cases for 
next-generation network technologies, 
such as C-V2X ~\cite{veh5g} and 802.11p/ITS-G5~\cite{vehbeacons}. In spite of the important
differences among these technologies, they all share the goal of
providing more network capacity to vehicles and  drivers.

Whenever more capacity is needed, millimeter-wave (mmwave)
communications are an appealing option~\cite{choi2016millimeter}. On
the negative side, mmwave  frequencies suffer from harsh propagation conditions,
with severe attenuation and high blockage probability. Such severe 
shortcoming has been addressed mainly by two approaches. On the
one hand, 
mobile network operators plan to backup data transfers served by
mmwave links by pairing them with  lower-frequency links especially to maintain connectivity through low-frequency control channels, so as to cope
with the unpredictable changes in real-world scenarios and the high
sensitivity of mmwave to the presence of obstacles \cite{giordani-tutorial, song-control}. 
On the other hand,  
the design of {\em directional} antenna systems, where the available
power is concentrated on one or more {\em beams}, can significantly help
to mitigate the problem\cite{bai-channel}.  An important trade-off in mmwave communications therefore arises between the
directionality gain that can be achieved using beamforming and the
spatial coverage that can be offered.

This implies that the
performance of mmwave networks critically depends on the beam design,
i.e., the number, direction, and amplitude of the beams. Successful beam design requires knowledge about the location of
the user(s) to serve, which explains why the earliest and currently, the most mature,
mmwave applications target static or quasi-static
scenarios. 
In addition, due to the mobility of the vehicular users, communication in mmwaves is even more sensitive to high Doppler shift and delays in channel status feedbacks \cite{song-wcomm}, which further complicates the beam training  and design phases.

We address the above issues by
leveraging the fact
that, in urban environments, it is possible to acquire a great deal of
information about vehicular mobility {\em without} detecting it in
real time. 

 Consider, for example, the commonly observed situation that red
traffic  lights are associated with a higher vehicle density and
lower speeds, two factors that can improve the achievable 
system throughput. This is very valuable information, and is known {\em a
  priori}. By exploiting this readily available information as well as static information such as road topology, we can facilitate efficient beam management without the need to make
real-time decisions. 
Our high-level goal is to assess the performance of this approach,
i.e., using traffic signal state information to complement and replace
real-time mobility data. To our knowledge, this is the first work that optimizes beam design
and carefully models mmwave coverage in vehicular networks, leveraging
both road topology and traffic regulation signal information. In summary, we make the following main 
contributions:
\begin{itemize}
\item[{\em (i)}] we first propose low-complexity approximations of the channel and beamforming
  behavior, tailored using empirical results  obtained through real-world traces and show their excellent agreement with existing, more complex, models.  Importantly, the applicability of our contribution goes beyond the scenario we study in this work;

    \item[{\em (ii)}] we formulate  beam configuration as an
optimization problem, aiming at maximizing the quality of coverage of
vehicular users, i.e., the users' sum rate. The choice of this objective, over others such as fairness or energy efficiency, reflects real-world
operational scenarios where, 
 as mentioned, it is
foreseen that mmwave
links are paired with low-frequency ones to ensure seamless connectivity. Thus, a sensible choice
is to maximize not the number of  mmwave links, but the number of
high-quality ones;

\item[{\em (iii)}] in light of the problem complexity, we define three
alternative beam design strategies, among which we
propose a low-complexity scheme, named Traffic Light (TL). Our scheme 
leverages  road topology information and traffic light signals to
efficiently configure the beams of the mmwave base stations;

    \item[{\em (iv)}] we show that TL closely matches the optimum,  in terms of sum rate, in a
      small-scale scenario. Then we evaluate all of the above strategies using  the large-scale
trace representing the real-world  topology and vehicular
mobility of Luxembourg City.  Our results show that, notwithstanding its much lower complexity, TL yields a performance
comparable to that of solutions based on real-time mobility
information.
\end{itemize}

In the following, we first discuss related work in \Sec{relwork} and
detail the proposed system and mmwave communication
models in \Sec{model}. Through numerical
simulations and real-world vehicular traces, \Sec{gain-model}  presents an accurate
and scalable
statistical model for the channel gain. 

\Sec{strategies} introduces  the proposed optimization problem and 
 the beam management strategies we envision, while   
\Sec{results} shows some results. Finally, \Sec{conclusion} concludes the paper.
 
\section{Related Work\label{sec:relwork}}

Frameworks on how to model and analyze mmwave cellular
networks have been provided in 
\cite{andrews-mmwave,akdeniz-mmwave},
while channel
characterization and modeling can be found in
\cite{rappaport-chanmodels, 3gppchanmodel}.

These
are statistical spatial models, based on experimental
measurements, that accurately describe the mmwave channel,  at a
cost of a high complexity. Generalized, simpler models for the channel
small-scale coefficients, as well as beamforming gains, can be found,
e.g., in \cite{bai-channel, andrews-mmwave}. Therein, the authors
assume that small-scale channel gains are subject to  Nakagami fading (i.e., a general model,  encompassing Rayleigh distribution,  that better suits mmwave propagation), while beamforming gains are modeled as discrete
random variables with a probability mass function determined by the
beamwidths of the transmitter and receiver beams.  

More related to our study is the recent work in \cite{zorzi-channel},
which approximates the
composite effects of the channel and beamforming gains, obtained under
the models in \cite{rappaport-chanmodels}, while also
considering realistic antenna radiation patterns. Compared to this work, we additionally
obtain the approximation of the channel and beamforming gains under
the channel model adopted by 3GPP \cite{3gppchanmodel} and, more
importantly, we differentiate between several levels of
beam misalignment.

Beam training is another important challenge in mmwave-based mobile networks, including vehicular ones, mainly due to
the associated control overhead and
delay~\cite{choi2016millimeter,giordani2017millimeter}. In \cite{wei-ia}, the authors  look at
exhaustive and iterative search procedures, and further propose a
two-stage hybrid training technique, in which the base station trains
in the first stage and the user equipment performs reverse training in
the second stage. In \cite{perfecto-v2v}, the authors propose a
framework that combines matching theory and swarm intelligence to
dynamically and efficiently perform user association and beam
alignment in a vehicle-to-vehicle (V2V) network. 

Other popular
approaches, more similar to the approach taken in this work,  are predicated on hoarding and leveraging as much location
information as possible, coming from the road-side sensors and 
the vehicles themselves~\cite{choi2016millimeter,discover-5g,zorzi-ia}, as well
as from out-of-band direction interference \cite{blind-steering}. 
In particular, using information originating from infrastructure nodes
is the key to beam design and switching in
\cite{iccbeamswitching,arxivali2019}. 
An inverse multipath fingerprinting approach is instead proposed in
\cite{inversefingerprint}, with the aim to match potential
beam directions with a vehicle location. 
Since collecting the information, processing
it, and re-aligning the beams in near-real-time is a very challenging
task, some works, e.g.,~\cite{loch2017mm}, envision
dispensing with beam realignment altogether, statically setting the
beam orientation using road topology information.

{\bf Novelty}.  
In addition to optimizing beam design, this study also assesses the viability and performance of TL, the low complexity scheme we propose. Importantly this study compares TL to the optimal configuration and other beam design strategies.
To our knowledge, this is the first
  work that uses a combination of road topology and traffic signal
  information, both of which can be obtained beforehand, to enable and
  facilitate beam management decision-making in vehicular
  networks.  
It should be noted that all of the above mentioned works focus on beam aligning for each
 gNB-user pair, implicitly assuming that each narrow beam is
employed to transmit to a single user only. However, in ultra dense
scenarios, it is highly likely that even a beam as narrow as
$5^{\circ}$ can cover several users simultaneously, 
users that can be multiplexed within the same beam. Thus, instead of focusing on perfectly aligning beams for each
gNB-user 
pair, we  look at how to align the beams at the gNB so as to
maximize the quality of   coverage of vehicles, i.e., the
users' sum rate. 
 In addition, we argue that we can achieve such solutions while
minimizing the use of real-time information, by utilizing instead, readily available, permanent, and deterministic information
such as road topology and traffic light signals.

Finally, we would like to mention that a preliminary version of this work has been presented in our
conference paper  \cite{noi-wowmom19}.

\section{Mmwave Vehicular Communication System\label{sec:model}}
In this section, we introduce the model
of the vehicular network and  we provide details on the physical layer model that we
adopt.   
We describe the signals transmitted by the mmwave base stations
(hereinafter referred to as gNBs)  and received by the
vehicle, and we 
specify how beams are generated by the
antennas.

\begin{table} 
\caption{Notation used in the system and communication model
    \label{tab:notation1}
} 
\scriptsize
\begin{tabularx}{\columnwidth}{|l|l|X|}
\hline
{\bf Symbol} & {\bf Type} & {\bf Meaning} \\
\hline\hline
\hline
$\Gc$ & set & Set of gNBs\\
\hline
$\Bc$ & set & Set of beams; the generic element is denoted by $b$\\
\hline
$\Vc$ & set & Set of vehicles; the generic element is denoted by $v$\\
\hline
$\Kc$ & set & Set of time intervals; the generic element is denoted by
              $k$\\
\hline
$P_{tot}(g)$ & parameter & Total power budget for gNB~$g$\\  
\hline
$N(g)$ & parameter & Maximum number of beams for gNB~$g$\\
\hline
$N_t$ ($N_r$) & parameter & Number of antenna elements per beam, at the transmitter
                            (receiver) side \\
\hline
$\Hm_c(g,v,k)$ & parameter & Channel matrix\\ 
\hline
$\dv(g,k,v)$ & parameter & Azimuth and elevation of vehicle~$v$ at
                           time~$k$ as observed by gNB~$g$ \\
\hline
$\psiv^{(g)}$ & parameter & Azimuth and elevation of the antenna array orientation
                at gNB $g$ \\
\hline
$\psiv^{(v)}(k))$ &  continuous variable & Azimuth and elevation of
                                           the antenna array
                                           orientation at vehicle~$v$ at
                           time~$k$\\
\hline
$\varphiv^{(g)}(b,k), \varphiv^{(v)}(b,k) $ &  continuous variable &
                                                                      $\deltav^{(g)}(b,k)  - \psiv^{(g)}$, \\
               & &                            $\deltav^{(v)}(b,k)-\psiv^{(v)}(k)$  \\
\hline
$\deltav^{(g)}(b,k)$, $\deltav^{(v)}(b,k)$ & continuous variable &
                                                                    Azimuth
                                                                    and
                                                                    elevation
                                                                    of
                                                                    beam~$b$
                                                                    at
                                                                    time
                                                                    $k$
                                                                    at gNB $g$ (vehicle $v$)\\ 
\hline
\end{tabularx}
\end{table}

 The notations we use throughout the paper are summarized in
Table\,\ref{tab:notation1}; 
the dependency of the parameters
and variables on the specific gNB, beam, or time step are omitted
whenever not needed.  
Also, in the following,  boldface lowercase letters denote column vectors
while  boldface uppercase letters denote matrices. The identity matrix is
represented by $\Id$. The transpose and the conjugate transpose of the
generic matrix $\Am$ are denoted, respectively, by $\Am\Tran$ and
$\Am\Herm$.

\begin{figure}
\centering
\includegraphics[width=0.4\textwidth]{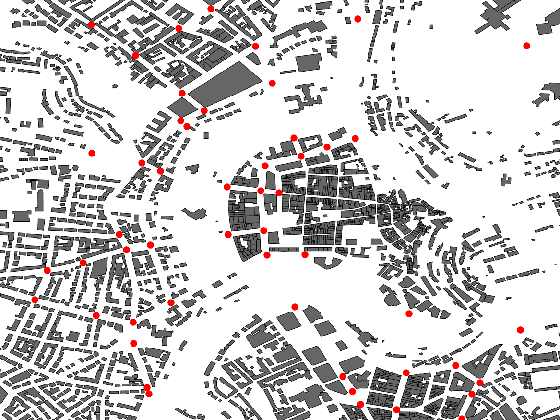}
\caption{\label{fig:scenario} Real-world scenario: Luxembourg city
  center. The red circles represent the locations of the traffic
  lights, on top of which the gNBs in our scenario are placed.}
\end{figure}

\subsection{Mmwave vehicular network: Model and data trace}
\label{subsec:system-trace}
Our network system  includes three main entities: 
gNBs~$g\in\Gc$, beams~$b\in\Bc$, and
vehicles~$v\in\Vc$. Also, time is divided in discrete
intervals~$k\in\Kc$.  
 To add a dose of realism to our problem formulation, we consider the scenario representing the real-world topology of
Luxembourg City, Luxembourg, and the  realistic mobility therein, as
reported  by the publicly-available, large-scale trace described in
\cite{lust}.  Luxembourg city is chosen partly due to the availability of open-source trace data, but also because its urban layout represents a typical European city.
The trace includes the location of traffic lights and bus stops, and 
the  traffic lights states (red, yellow, green) at all time steps. It
represents  around 14,000 vehicles over a period of 12
hours, generated with SUMO, an open source multi-modal traffic simulator, and based on real-world traffic flows,
e.g., commuters traveling to the city center.   
Depicted in \Fig{scenario} is the  $2$\,km\,$\times 2$\,km area of the city center we take into consideration. Throughout this road topology, we place
a total of 51~gNBs, colocated to traffic lights (their positions are
marked by red dots in \Fig{scenario}); the traffic lights corresponding to
intersections with highest vehicle density were selected.

The gNBs and the vehicles are equipped with uniform planar array
(UPA) antennas, composed of a grid of antenna elements 
spaced by $\lambda/2$. 
The vehicle UPA is only capable of analog beamforming
  (it is equipped with a single radio frequency (RF) chain). 
Digital beamforming is instead enabled at the gNBs, with $N(g)$ being
the maximum number of beams that can be simultaneously active at the
generic gNB $g$. We also denote by $P_{tot}(g)$ the total
power budget the gNB has to split among its beams.  
As for the antenna elements, we consider two different models: 
\begin{itemize}
\item[{\em a)}] the isotropic radiation antenna element denoted as ``ISO'';
\item[{\em b)}] the 3GPP sectored antenna element 
  \cite{3gppchanmodel}, denoted by ``3GPP'', which is characterized
  by an 8\,dBi maximum directional gain, significantly smaller side
  lobes than the ISO element, and higher front-to-back ratio. In this case, a three-sector
  gNB implementation is assumed, and the corresponding beamforming
  vector is used to shift the beam within each $120^{\circ}$ sector.
\end{itemize}

In this scenario, the goal is to make beamforming decisions that maximize the aggregated
data rate over the vehicular users.  We focus on downlink
communications, although the analysis can easily be
  extended\footnote{Since the vehicles are equipped with a single RF
    chain, and can transmit on a single beam at a time, beam
    management in uplink is indeed straightforward.}.

Such decisions concern: 
\begin{itemize}
\item[(i)] the number of active
beams at each gNB, 
\item[(ii)] their width and direction, and 
\item[(iii)] the transmission power assigned to each beam. 
\end{itemize}

The decisions are made at the gNBs every time step $k$, which coincides with the channel coherence time\footnote{We recall
that channel estimation is performed through 
 periodically transmitted  pilot symbols (e.g., CSI Reference
 Symbols (CSI-RS) in 5G New Radio (NR)). }.  For simplicity, we assume perfect channel state information (CSI) knowledge at the gNB. We consider that a beamshifting vector is used at the gNB to point the beam towards the desired direction, as described in Sec.~\ref{subsec:channel-model}, while the  width of the beam can be selected from a limited set of beamwidths, which can be obtained by varying the number of antenna elements employed.

The gNB transmits a
single data stream towards one or more vehicles. 
As envisioned in 5G NR, data transmissions are organized into time slots, during which the channel
is assumed to be constant. Thus, we can consider 
that, during data transfers, pilot symbols (the so-called DeModulation Reference Symbols (DMRS))
are transmitted in  every time
slot to adapt the modulation scheme to the instantaneous channel conditions.
Below, we will denote by $C$ the number of slots per time step, and by $S$ the number of symbols per
slot.

\subsection{Mmwave communication model\label{subsec:channel-model}}

\begin{figure}[t]
  \centerline{\resizebox{1.0\columnwidth}{!}{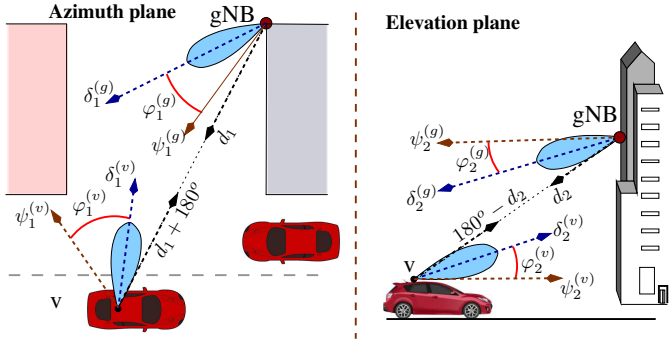}}
  \caption{mmwave communication beams and notations used.}
  \label{fig:angles}
\end{figure}

Given a gNB, $g$, and consider that the orientation in space of its antenna elements
 is characterized by the azimuth-elevation pair of
angles $\psiv^{(g)}=[\psi^{(g)}_1,\psi^{(g)}_2]$, as depicted in Fig.~\ref{fig:angles}. 

The orientation of the antenna elements at vehicle $v$ is
represented by the pair of angles
$\psiv^{(v)}=[\psi^{(v)}_1,\psi^{(v)}_2]$, while the azimuth-elevation
angle of the vehicle $v$, as observed by the gNB $g$ during time step
$k$, is denoted by $\dv=[d_1,d_2]$. 

At every time step $k$, 
the gNB adjusts the antenna gains to generate one or more
beams.  The number of beams to be generated is decided by the beam design algorithm. Let us focus on the generic beam $b$, with azimuth-elevation
angle  $\deltav^{(g)}$.  
To generate the beam, the gNB employs the beamforming vector
$\vv(\varphiv^{(g)})$ of size $N_t=N_{t,1}\times N_{t,2}$, with $n$-th element: 
\begin{equation} [\vv(\varphiv^{(g)})]_n =
  \frac{1}{\sqrt{N_t}}\ee^{\jj \pi (n_1\cos\varphi^{(g)}_2\sin \varphi^{(g)}_1+n_2\sin
    \varphi^{(g)}_2)}\end{equation} where the integers
$n_1\in\{1,\ldots, N_{t,1}\}$ and $n_2\in\{1,\ldots, N_{t,2}\}$ are
such that $n=n_1N_{t,2}+n_2$.  The vector of angles
$\varphiv^{(g)}=\deltav^{(g)}-\psiv^{(g)}$ represents the direction of
the beam with respect to the transmitting antenna orientation
$\psiv^{(g)}$. If $\deltav^{(g)}=\dv$, then the beam of the transmitter
is perfectly aligned with the vehicle on both the azimuth and
elevation planes.

The symbols transmitted by the gNB in the $c$-th slot, $c=1,\ldots,C$, are represented by the
vector $\xv_c=[x_{c,1},\ldots,x_{c,S}]\Tran$ whose elements
are modeled as independent, complex random variables with zero-mean and
variance $\EE[|x_{c,s}|^2] = P$ ($s=1,\ldots,S$) where $P$ is the power of the considered beam. 
In the following, for  
simplicity, we omit the argument $\varphiv^{(g)}$ of the beamforming
vector $\vv$.
The signal transmitted by the gNB in the $c$-th  slot  
is represented by the $N_t\times S$ matrix $\vv\xv_c\Tran$ and
the corresponding signal received by the vehicle is described by the
$N_r\times S$ matrix $\Ym$, with $N_r=N_{r,1}\times N_{r,2}$: 
\begin{equation}\label{eq:Y}
  \Ym_c = \Hm_c\vv\xv_c\Tran + \Em_c \,.
\end{equation}
In (\ref{eq:Y}), $\Hm_c$ is the $N_r\times N_t$ channel matrix between
 gNB $g$ and  
vehicle $v$ during the transmission on the  $c$-th slot,  while $\Em_c$ is a
term accounting for noise and interference. Note that  
nearby gNBs cause interference if the radiation pattern of their beams 
affects the vehicle reception.

In a typical mmwave urban scenario, the channel between the
transmitter and the receiver is composed of a  number of path
clusters, each cluster corresponding to a macro-level scattering
path. The $\ell$-th path is described by its amplitude, $h_{\ell,c}$,
and the azimuth-elevation departure and arrival pairs of angles
$\varthetav^{(g)}_{\ell,c}$ and $\varthetav^{(v)}_{\ell,c}$, measured
with respect to the antenna orientation angles $\psiv^{(g)}$ and
$\psiv^{(v)}$, respectively. Using the model provided in
\cite{rappaport-chanmodels, akdeniz-mmwave, 3gppchanmodel}, the channel matrix
during the $c$-th slot, $\Hm_c$, can be modeled as
\begin{equation}\label{eq:H} \Hm_c = \sqrt{\frac{1}{L}}\sum_{\ell=1}^L h_{\ell,c}
  \av_v(\varthetav^{(v)}_{\ell,c})\av_g(\varthetav^{(g)}_{\ell,c})\Herm \end{equation}
where $L$ is the number of paths, and  $\av_g(\varthetav)$ and
$\av_v(\varthetav)$ are 
the  response vector for, respectively, the transmit and the receive antenna.
More specifically, the $n$-th element of $\av_g(\varthetav)$ is:
\begin{equation}
  [\av_g(\varthetav)]_n = \ee^{\jj \pi (n_1\cos\vartheta_2\sin \vartheta_1+n_2\sin \vartheta_2)}
\end{equation}
where $n_1\in\{1,\ldots, N_{t,1}\}$ and $n_2\in\{1,\ldots, N_{t,2}\}$ are such that $n=n_1N_{t,2}+n_2$.
Similarly, the  $n$-th element of $\av_v(\varthetav)$ is given by
\begin{equation}
  [\av_v(\varthetav)]_n = \ee^{\jj \pi  (n_1\cos \vartheta_2\sin \vartheta_1+n_2\sin \vartheta_2)}
\end{equation}
where $n_1\in\{1,\ldots, N_{r,1}\}$ and $n_2\in\{1,\ldots, N_{r,2}\}$
are such that $n=n_1N_{r,2}+n_2$.  The $N_r\times S$ matrix $\Em_c$
in~\eqref{eq:Y} has complex Gaussian independent entries, and its
columns $\ev_{c,s}$ have zero mean and covariance
$\EE[\ev_{c,s}\ev_{c,s}\Herm]=I_c\Id$ ($s=1,\ldots,S$).  Since the vehicle UPA
is only capable of analog beamforming, we consider that 
the weight vector $\wv(\varphiv^{(v)})$ of size $N_r$ is
applied to the RF chain, with generic element: 
  \begin{equation}\label{eq:w}
    [\wv(\varphiv^{(v)})]_n = \ee^{\jj\pi (n_1 \cos \varphi^{(v)}_2\sin \varphi^{(v)}_1
      +n_2 \sin \varphi^{(v)}_2)} \,,
  \end{equation}
  in order to generate a beam with azimuth-elevation angle 
  $\deltav^{(v)}=\varphiv^{(v)}+\psiv^{(v)}$. The integers
  $n_1\in\{1,\ldots, N_{r,1}\}$ and $n_2\in\{1,\ldots, N_{r,2}\}$ are
  such that $n=n_1N_{r,2}+n_2$. This scenario is illustrated in Fig.~\ref{fig:angles}.
  If $\delta^{(v)}_1={\rm mod}(180^\circ+d_1,360)$ and $\delta^{(v)}_2=180-d_2$,  
  then the receiver's beam  is perfectly aligned with the gNB, in azimuth and elevation.
  
Through the weighting procedure, the receiver forms the vector 

    $\zv_c = \wv\Herm\Ym_c= \widetilde{h}_c \xv_c\Tran + \wv\Herm\Em_c$

    where
    \begin{equation} \label{eq:tildeh}
        \widetilde{h}_c = \wv\Herm\Hm_c\vv
      \end{equation}
      is the scalar channel coefficient summarizing the effects of the signal
      propagation conditions and of the antenna and beam design. For
      simplicity,  in
    the above equations,  we omitted the argument
      $\varphiv^{(v)}$ of $\wv$. Finally, estimates of the symbols belonging to the $c$-th slot
      are obtained at the receiver by processing the vector $\zv_c$.

Note that, if more than one gNB transmits toward the same user  in such
a way that it causes  constructive interference  (CoMP-like mode), similar
expressions to the ones above hold, where $\Hm_c \vv$  represents the
equivalent channel.

\section{A Convenient Statistical Model of the Channel Gain\label{sec:gain-model}}

In this section, we provide a {\em statistical characterization} of the
channel gain under different antenna models and beam alignment
conditions, and we show its validity in the real-world scenario
represented by the Luxembourg trace,  depicted in \Fig{scenario}. Note, however, 
that our model has general validity and can be
used for the study of aspects beyond the scope of this work.

We start by  applying two channel models, among those which appeared in
the literature, that are most commonly used: 
\begin{itemize}
  \item the ``NYU'' model \cite{akdeniz-mmwave}, a statistical spatial
channel model based on experimental measurements performed in the city
of New York; 
\item the ``3GPP'' model \cite{3gppchanmodel}, a
hybrid of geometry-based stochastic and map-based channel models,
which has been adopted by 3GPP. 
\end{itemize}
Both these models describe the channel between two communicating
endpoints as composed of several clusters of paths, each synthesized
using a large number of subpaths, with certain azimuth and elevation
arrival and departure angle. 

In the following, we implement both the above  models and, using
link-level simulations, we provide a statistical characterization of
the channel gain $\widetilde{h}_c$ as defined in~\eqref{eq:tildeh}.

\begin{figure*}[t]
  \centerline{\resizebox{0.9\textwidth}{!}{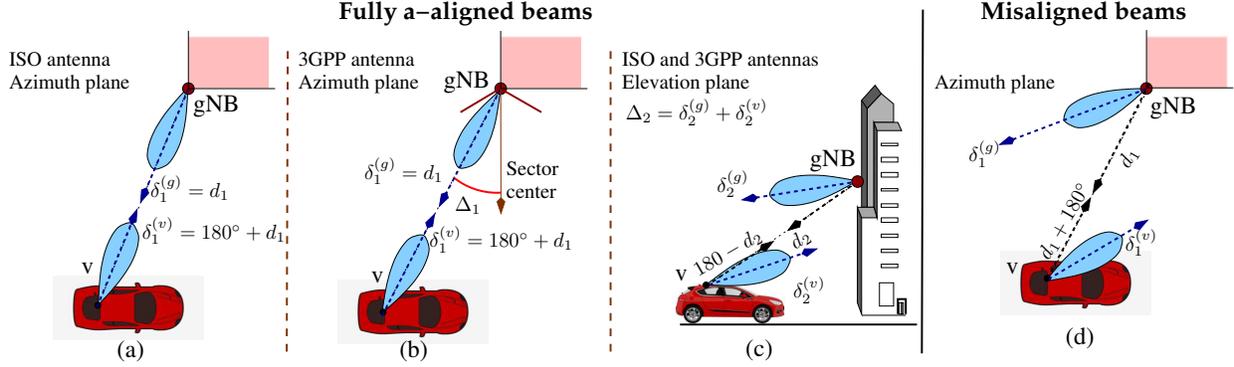}}
  \caption{The case of full alignment along the Azimuth ((a), (b), and
    (c))
    cases), and the case of misaligned beams (d).}
  \label{fig:beams}
\end{figure*}

Given the channel matrix $\Hm_c$ given in \Eq{H}, the beam directions (specified
by the weight vectors $\vv$ and $\wv$), and the relative positions
between  transmitter and  receiver,  we characterize the
gain of the  channel connecting  gNB $g$ to  vehicle $v$ during the
transmission in the $c$-th slot. To this end, we rewrite
(\ref{eq:tildeh}) as:
\begin{equation}\label{eq:gain}
  |\widetilde{h}_c|^2=a G
\end{equation}
where $a$ is the path loss between the gNB-vehicle pair as a
function of their distance, and the gain component, $G$,
encompasses both the small-scale fading effects of the channel and 
 the beamforming gain of the antenna array.  

The value of $G$ has a strong dependence on the degree of alignment between the beams; under line of sight (LoS) conditions, we have  three cases:
\begin{itemize}
\item ``fully a-aligned'' link, where both beams are
  perfectly aligned with each other along the azimuth, 
\item ``partially a-aligned'' link where the beams are aligned
  along the azimuth at only one end, at  either the
  transmitter or the receiver side, and 
\item ``misaligned'' link where neither beam is aligned.
\end{itemize}

The accurate description of the behaviour of gain $G$ in these three cases is critical  in
modeling the useful power and interference levels at the receiver
side, especially in the presence of LoS. 
In particular, we differentiate between the 
partially a-aligned and fully misaligned cases,  since simulations show that the gain can
be significantly higher when the beam  is aligned at one end, as
opposed to when it is misaligned at both ends.  The partially
a-aligned gain can be applied to model more accurately  the
interference experienced at the receiver, especially  in those cases
where a vehicle may be inadvertently under
the coverage of an interfering gNB, or to describe  the constructive interference  in the context of CoMP-like cases.

In the following subsections, we empirically characterize the distribution of the
gain $G$ under the above described propagation conditions and for the two different antenna
radiation patterns. For each scenario, we show that the empirical distribution of
$G$, 
can be tightly
approximated by using Gaussian, exponential, or log-logistic distributions.

\subsection{Fully a-aligned beams and 3GPP channel}\label{subsec:fullyaligned3gpp}
When LoS conditions hold and the beams of the transmitter and of the receiver are perfectly
aligned in the azimuth plane (a-aligned), we found that, under  the ``3GPP''
channel model, the distribution of the
gain $G$ is well  approximated by a Gaussian distribution
with mean $\mu_G$ and variance $\sigma_G^2$.

Specifically, when ISO antenna elements are employed, the mean
$\mu_G$ is given by 
\begin{equation}\label{eq:mug}
  \mu_G(\Delta_2)=\mu_{0}\exp\left(-\frac{\Delta_2^2}{\gamma_{\mu}^2}\right)\,.
\end{equation}
Since we
do not consider perfect alignment on the elevation plane, our analysis
shows that the parameter $\mu_G$ depends on the angle  of misalignement in the 
elevation plane, denoted by $\Delta_2$ (see Fig.~\ref{fig:beams}-(c) for details).
When instead 3GPP antenna elements are used, $G$ also depends on the angle
$\Delta_1$ between the direction of the beam at the gNB and the direction of the sector
center\footnote{Sector antennas with 3GPP antenna elements are
  implemented only at the gNB, while at the vehicle we only consider
  isotropic antennas.} (see Fig.~\ref{fig:beams}-(b)).
In particular, the resulting gain is reduced when the beam at the gNB
is not aligned with  the direction of the sector, and it becomes 0 for a
beam out of the sector width, as per the 3GPP antenna element
radiation pattern \cite{3gppchanmodel}. The  mean
$\mu_G$, for the 3GPP antenna element, is given by: 
\begin{equation}\label{eq:mug2}
  \mu_G(\Delta_1,\Delta_2)=\mu_0\exp\left(-\frac{\Delta_2^2}{\gamma_{\mu}^2}\right)10^{-1.2\left(\frac{\Delta_1}{\theta_{\rm 3dB}}\right)^2}
\end{equation}
where $\theta_{\rm 3dB}=65^{\circ}$ is the half-power beamwidth of a single antenna element. 
With regard to the standard deviation $\sigma_G$, in both the ISO and
3GPP cases, it  
      depends on $\Delta_2$ according to the  relation:  
\begin{equation}\label{eq:sigmag}
  \sigma_G(\Delta_2)=\sigma_{0}\exp\left(-\frac{\Delta_2^2}{\gamma_{\sigma}^2}\right) \,.
\end{equation}
Finally, we established a dependence between the parameters $\mu_o, \sigma_0,
\gamma_\mu,\gamma_g$ and the product  $N_tN_r$, captured by the expressions provided in Table \ref{table:3gppaligned}.

\begin{table} 
   \caption{Values of the parameters characterizing the distribution
     of $G$, for fully a-aligned beams and 3GPP  channel model}
   \begin{center}
 \begin{tabular}{| l | l | l |}
 \hline
 & ISO & 3GPP \\ 
\hline
 $\mu_0$&$0.537(N_tN_r)^{0.998}$&$3.26(N_tN_r)$ \\ \hline
 $\sigma_0$&$0.23(N_tN_r)^{0.7}$&$1.33(N_tN_r)^{0.65}$\\ \hline
 $\gamma_{\mu}$&$55.02(N_tN_r)^{-0.287}$&$49.01(N_tN_r)^{-0.274}$ \\ \hline
 $\gamma_{\sigma}$&$55.86(N_tN_r)^{-0.28}$&$55.86(N_tN_r)^{-0.28}$ \\ \hline
 \end{tabular}
\end{center}
\label{table:3gppaligned}
 \end{table}

\begin{figure} 
\includegraphics[width=0.45\textwidth]{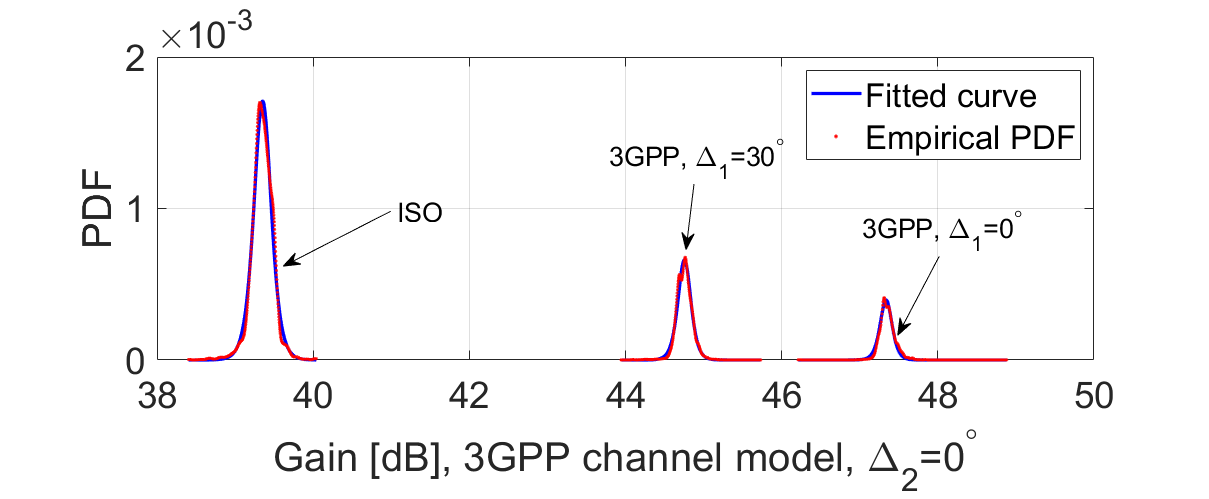}
\centering
\includegraphics[width=0.45\textwidth]{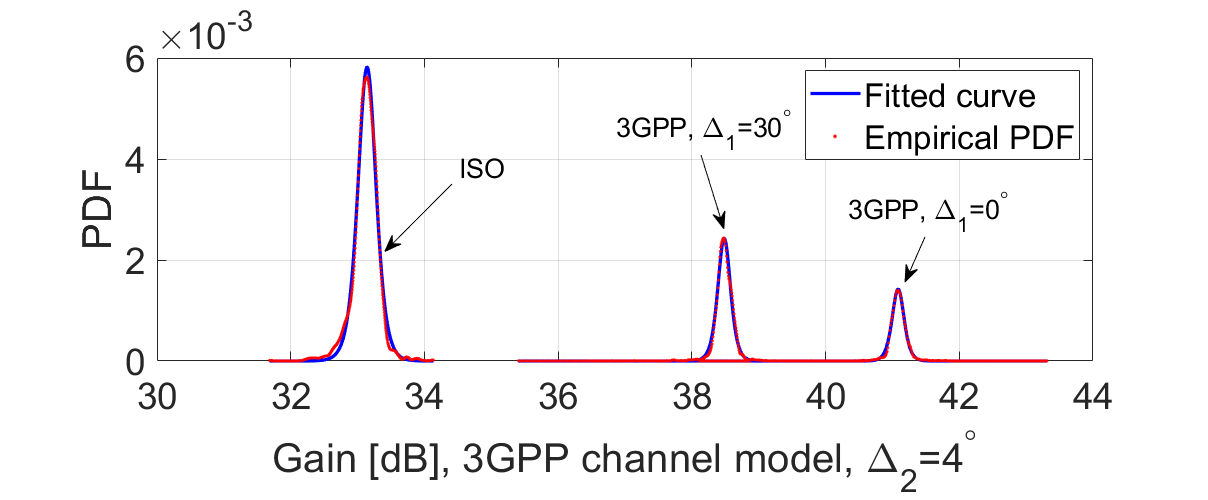}
\caption{\label{fig:3gpp-alignfit}  LoS and fully a-aligned case: Empirical and fitted
  distribution of $G$ under the 3GPP
  channel model, and when the ISO and 3GPP antennas are employed. $\Delta_2=0^{\circ}$ (top) and $\Delta_2=4^{\circ}$ (bottom). $N_t=256$, $N_r=64$.}
\end{figure}

 Fig.~\ref{fig:3gpp-alignfit} 
compares the empirical\footnote{The fitting was performed using
the  MATLAB curve fitting toolbox.}
 distribution of the gain $G$ to the Gaussian distribution whose
 parameters are specified in~\eqref{eq:mug}--\eqref{eq:sigmag}.   The plots {of the probability density function (PDF) of the gain} have been obtained under the 3GPP
 channel model, for both ISO  and 
3GPP antenna model. Importantly, our proposed
approximation is very tight in both cases; furthermore, it 
 holds also for
 other values of the number of antennas, $\Delta_2$, and 
$\Delta_1$ (for the 3GPP antenna model).

\begin{figure} 
\centering
\includegraphics[width=0.22\textwidth]{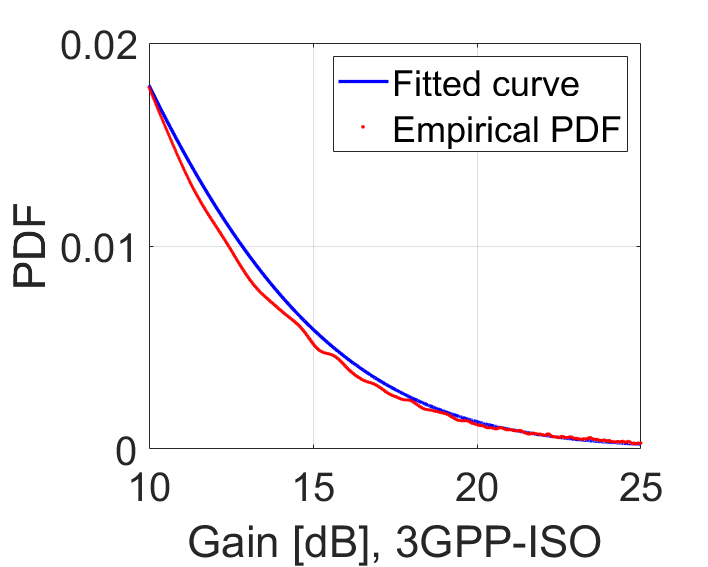}
\includegraphics[width=0.22\textwidth]{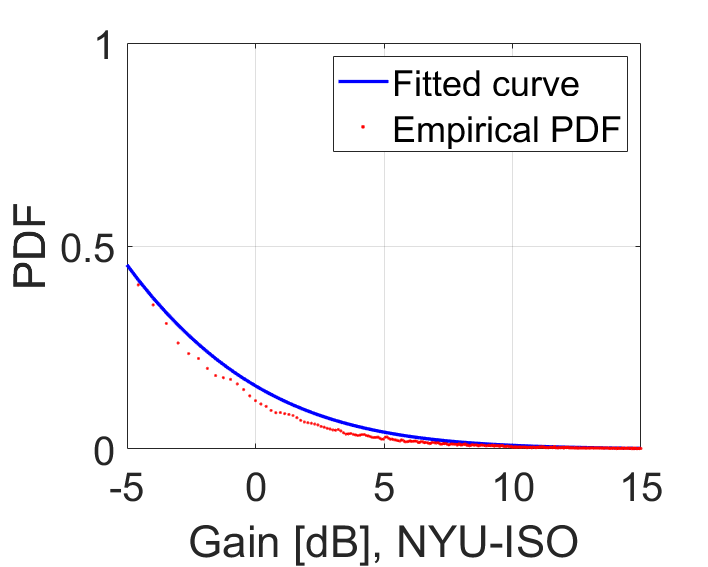}
\caption{\label{fig:nlosfit}  NLoS conditions: Empirical and fitted
  distribution of gain $G$ under the 3GPP channel model (left) and
  the NYU channel model (right), when ISO antennas are employed. $N_t=256$, $N_r=64$.}
\end{figure}

To be able to compare to the no line of sight (NLoS)
  case, we derived the results shown in Fig.~\ref{fig:nlosfit}
(left). In NLoS conditions, the
  values of $G$ drop significantly and  the gain distribution 
  can be better approximated with the {\em{log-logistic}}
  distribution: 
 \begin{equation} \label{eq:loglogdist}
  f_{G}(x)=\frac{1}{mx}\frac{e^z}{(1+e^z)^2}
    \end{equation}
where $z=(\log(x)-m)/s$. 

Due to the absence of the LoS path, no significant dependency on the misalignment elevation
angle $\Delta_2$ could be inferred, nor a straightforward relationship
to the product $N_tN_r$.  However,  as shown in Table
\ref{table:nlosparams}, the values of the distribution parameters $m$
and $s$  vary depending on the channel model/antenna type considered,
and the number of antennas $N_t$ and $N_r$. 
Due to space limitations, in Fig.~\ref{fig:nlosfit}, we only show the results for the ISO antenna elements, however the approximations hold  as well when 3GPP antenna elements are used. 
  
\begin{table} 
  \caption{Values of the parameters characterizing the distribution of $G$, when the beams are a-aligned in NLoS}
 \begin{tabularx}{\columnwidth}{|X|| X | X |X|}
 
 \hline
 (m,s)& $N_t=256$ &$N_t=64$ &$N_t=64$  \\ 
& $N_r=64$ & $N_r=64$ & $N_r=16$\\
\hline
\multirow{1}*{3GPP-ISO} &$(2.97, 0.99)$&$(3.83, 0.98)$&$(3.28, 0.97)$\\   \hline 
\multirow{1}*{3GPP-3GPP}& $(4.63, 1)$&$(5.57, 0.99)$&$(4.98, 0.98)$\\
 \hline 
\multirow{1}*{NYU ISO}& $(-1.7, 0.97)$&$ (-1.98, 0.97)$&$(-2.75, 0.95)$\\  \hline 
\multirow{1}*{NYU 3GPP}& $(-2.6, 1.03)$&$ (-3.68, 2.72)$&$(-0.48, 1.1)$\\  \hline 
 \end{tabularx}
 \label{table:nlosparams}
 \end{table}

\subsection{Fully a-aligned beams and NYU channel}\label{subsec:fullyalignednyu}
\begin{figure*} 
\centering
\includegraphics[width=0.32\textwidth]{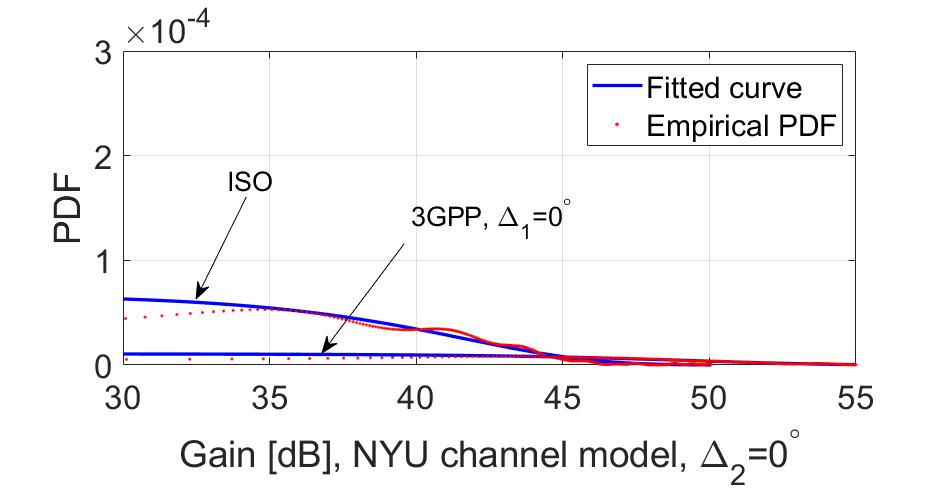}
\includegraphics[width=0.32\textwidth]{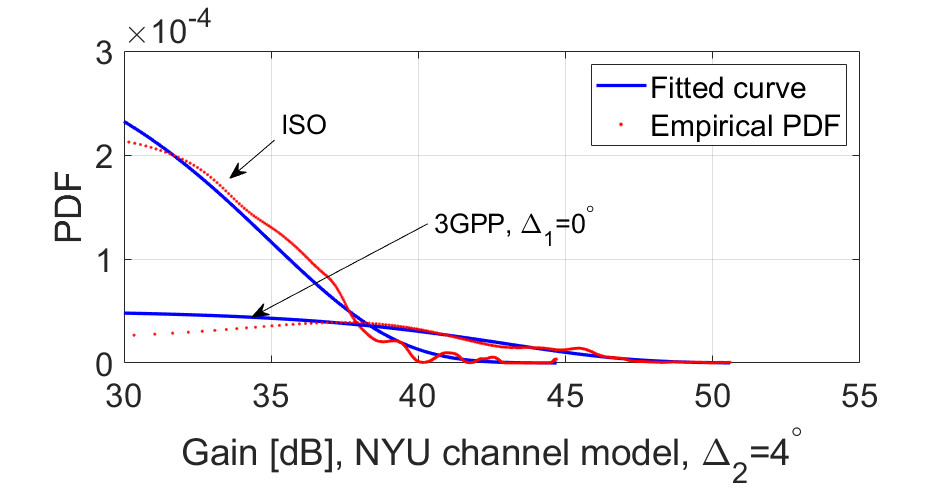}
\includegraphics[width=0.32\textwidth]{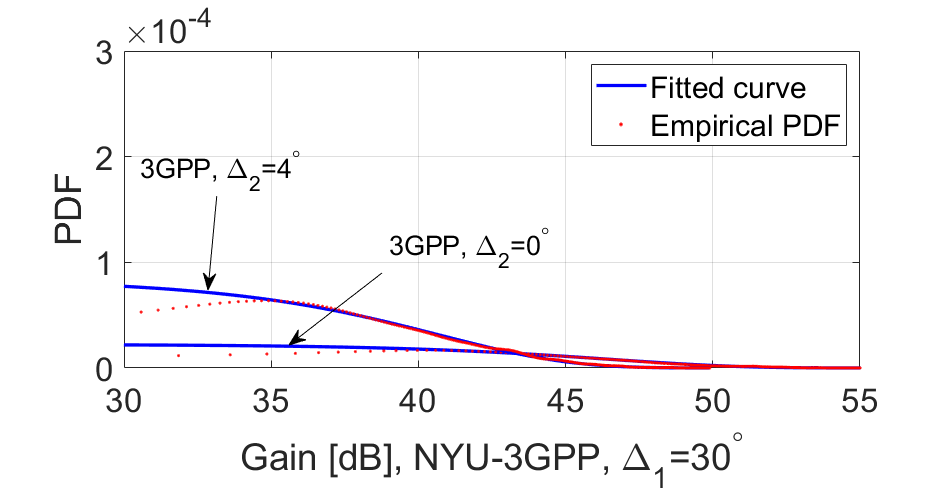}
\caption{\label{fig:nyu-alignfit} LoS and fully a-aligned case:    Empirical and fitted
  distribution of $G$ under the NYU channel 
  model and when ISO and 3GPP antennas are  
  employed. $\Delta_2=0^{\circ}$ (left), $\Delta_2=4^{\circ}$
  (middle), and $\Delta_1=30^{\circ}$ (right)  . $N_t=256$, $N_r=64$,
  and   $\Delta_1=0^{\circ}$ for 3GPP antennas.}
\end{figure*}

Now, we still consider  fully a-aligned beams, but we 
focus on the NYU channel model. In this case and under LoS conditions, we find that the empirical distribution of the gain $G$
is tightly approximated by an exponential
distribution with average $\alpha_G$, i.e.,
 $ f_G(x)=\frac{1}{\alpha_G}\exp\left(-\frac{x}{\alpha_G}\right)$.
In particular, when ISO antenna elements are employed, the parameter $\alpha_G$ depends on $\Delta_2$ according to 
  $\alpha_G(\Delta_2)=\alpha_0\exp\left(-\frac{\Delta_2^2}{\gamma_{\alpha}^2}\right)$.
Instead, with 3GPP antenna elements, $\alpha_G$ depends on both $\Delta_1$ and
$\Delta_2$, according to the relation: 
\begin{equation}
  \alpha_G(\Delta_1,\Delta_2)=\alpha_{0}\exp\left(-\frac{\Delta_2^2}{\gamma_{\alpha}^2}\right)10^{-1.2\left(\frac{\Delta_1}{\theta_{3dB}}\right)^2}
\end{equation}
where $\theta_{\rm 3dB}=65^{\circ}$.
Interestingly, the values of the parameters $\alpha_0$ and $\gamma_{\alpha}$ can be expressed as functions of the
product $N_tN_r$, as shown in Table \ref{table:nyualigned}.
 \begin{table} 
   \caption{Values of the parameters of the
     distribution of $G$, under LoS and  NYU channel model}
   \begin{center}
 \begin{tabular}{| l | l | l |}
 \hline
 & ISO & 3GPP \\ 
\hline
 $\alpha_0$&$0.63(N_tN_r)^{1.05}$&$5.2(N_tN_r)^{1.03}$ \\ \hline
 $\gamma_{\alpha}$&$54.85(N_tN_r)^{-0.3}$&$54.85(N_tN_r)^{-0.3}$ \\ \hline
 \end{tabular}
 \end{center}
 \label{table:nyualigned}
\end{table}
Examples of the empirical and fitted distribution of $G$ are provided
in Fig.~\ref{fig:nyu-alignfit} 
when the gNB is equipped with ISO and 3GPP antenna elements, and for different values of $\Delta_1$ and $\Delta_2$.
Again, there is a very close match between the experimental
distribution of the gain and our analytical approximation, which 
holds also for different values of numbers of antennas, $\Delta_2$,
and $\Delta_1$. 

Similarly to the 3GPP channel model, in the absence of LoS, the a-aligned gain for the NYU channel model  is better approximated using the log-logistic distribution, given by (\ref{eq:loglogdist}), however different values for parameters $m$ and $s$ are applied, as reported in Table \ref{table:nlosparams}. The empirical distribution and the fitted counterpart are shown in the right plot in Fig.~\ref{fig:nlosfit}.

  \begin{figure} 
 \includegraphics[width=0.22\textwidth]{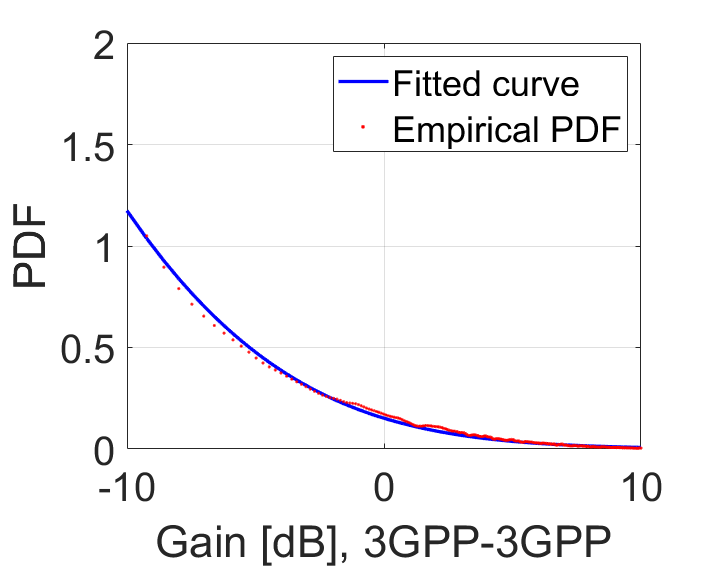}
\includegraphics[width=0.22\textwidth]{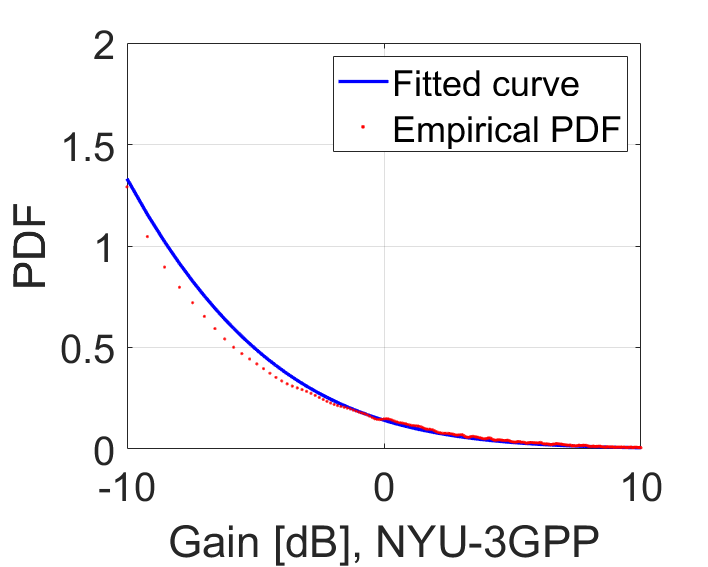}
\centering
\includegraphics[width=0.22\textwidth]{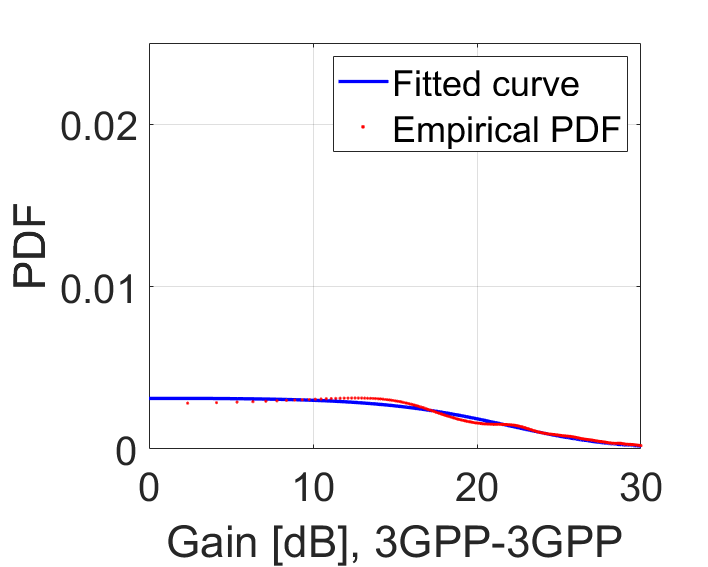}
\includegraphics[width=0.22\textwidth]{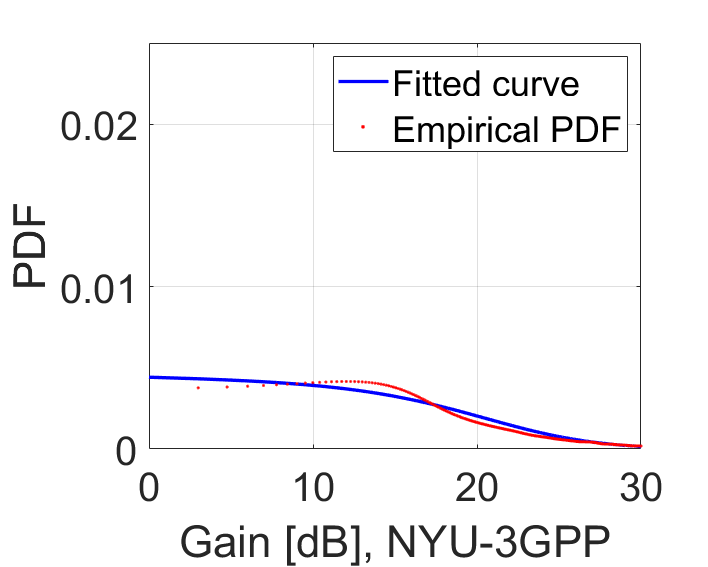}
\includegraphics[width=0.22\textwidth]{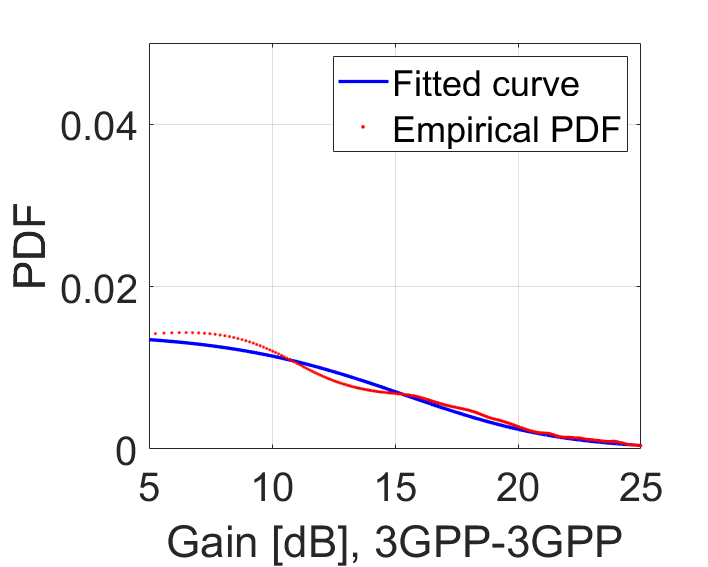}
\includegraphics[width=0.22\textwidth]{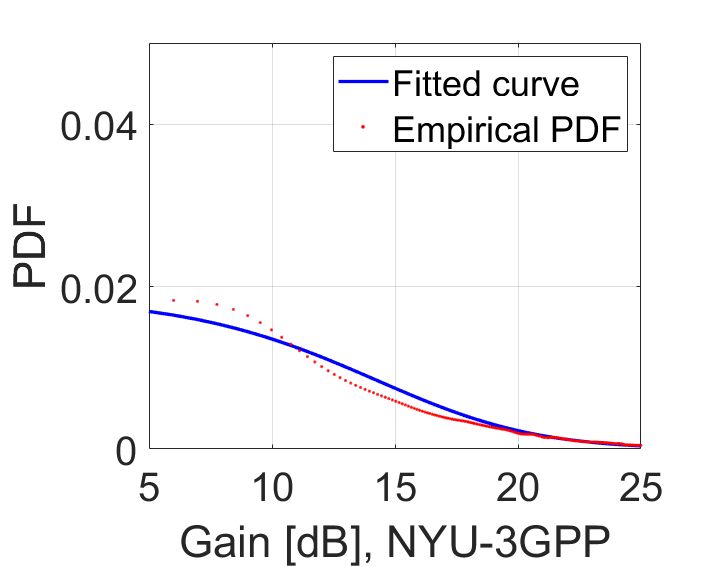}
\caption{\label{fig:misalignedatrx} Empirical and fitted
  distribution of gain $G$,  for the  3GPP (left) and
  the NYU (right) channel model when 3GPP antennas are employed, in the case of full misalignment (top), partial a-alignment at the transmitter (middle) and partial a-alignment at the receiver (bottom). $N_t=256$, $N_r=64$.}
\end{figure}

\subsection{Partially a-aligned or misaligned beams}\label{subsec:partiallyaligned}
When the beams are partially a-aligned or misaligned, we found that the empirical distribution of $G$ can be
approximated with the log-logistic distribution reported in (\ref{eq:loglogdist})

The approximation is valid for all considered channel models and 
antenna elements, however the values of the parameters $m$ and $s$ 
vary depending on the type of misalignment. Namely, we consider three cases of misalignment: 1) full misalignment where beams both at the transmitter and receiver are not aligned, 2) the partial a-alignment at the transmitter side, where the beam at the transmitter is aligned but not at the receiver, and 3) the partial a-alignment at the receivers side, where the beam at the receiver is aligned but not at the transmitter. 
  \begin{table} 
  \caption{Values of the parameters characterizing the distribution of $G$, when the beams are partially a-aligned or misaligned}
 \begin{tabularx}{\columnwidth}{|X|| X | X | X |} 
 \hline
 $(m,s)$& $N_t=256$ & $N_t=64$ &$N_t=64$ 
   \\ 
& $N_r=64$ & $N_r=64$ & $N_r=16$\\
\hline
\multirow{1}*{3GPP-ISO}& 1. $(-2.5, 1)$&$1. (-1.45, 1)$&$1. (-1.65, 1)$\\ \cline{2-4}
 &2.$(3.89, 0.99)$&$2. (3.2, 0.99)$&$2. (3.15, 0.85)$\\  \cline{2-4}
 &3.$(2.35, 0.98)$&$3. (3.2, 0.99)$&$3. (1.9, 0.98)$\\  \hline \hline
\multirow{1}*{3GPP-3GPP}& 1.$(-1.05, 1)$&1.$(-0.16, 1)$&1.$(0.01, 1)$\\ \cline{2-4}
& 2.$(5.72, 0.99)$&2.$(5.05, 0.99)$&2.$(4.99, 0.86)$\\ \cline{2-4}
& 3.$(4.16, 0.98)$&3.$(5.03, 0.99)$&3.$(3.79, 0.98)$\\ 
 \hline \hline
\multirow{1}*{NYU ISO}& 1.$(-1.96,1.01)$&1.$ (0.59, 1.01)$&1.$(-0.17,1.06)$\\ \cline{2-4}
& 2.$(3.77, 0.99)$&2.$ (3.3, 1)$&2.$(2.97, 0.98)$\\ \cline{2-4}
& 3.$(2.17, 0.98)$&3.$ (2.94, 0.99)$&3.$(1.8, 0.98)$\\ \hline \hline
\multirow{1}*{NYU 3GPP}&$1. (-1.48, 0.97)$&$1. (1.65, 0.99)$&$1. (-0.97, 1.05)$ \\\cline{2-4}
& 2.$(5.48, 0.99)$&2.$ (4.99, 1)$&2.$(4.76, 0.98)$\\ \cline{2-4}
& 3.$(3.95, 0.99)$&3.$ (4.75,0.99)$&3.$(3.51, 0.98)$\\   \hline 
 \end{tabularx}
 \label{table:misalignedparams}
 \end{table}

Unlike in the fully a-aligned case, in the partially a-aligned
case, no evident dependency could be established
between the misalignment angle in the azimuth plane and the values of
the distribution parameters $m$ and $s$. 
This is due to  the fact that the misalignment angle only
determines the contribution to the gain coming from the side-lobes of
the misaligned end, while the dominant contribution to the resulting
effective gain comes from the main lobe of the aligned end, which is
not affected by this angle. Hence, the dependency of the gain on the
misalignment angle tends to vanish and no characterization of it is possible. 
Similarly, although it is evident that the parameters of the
distribution $m$ and $s$ are affected by the number of antennas $N_t$
and $N_r$, no straightforward relationship can be established.  
The numerically evaluated values for the parameters $m$ and $s$ are provided in
Table \ref{table:misalignedparams}, for all possible combinations of
channel model and antenna elements (across rows) and for several values of
$N_t$ and $N_r$ (across columns).  In each cell of the table, the
three lines correspond, respectively, to misalignment at both transmitter and receiver, alignment at the transmitter only, and alignment
at the receiver only.

A close match can be observed between the empirical samples and the
fitted distributions, in
Fig.\,\ref{fig:misalignedatrx} for misaligned links (top),  
partially a-aligned links at the transmitter (middle) and partially a-aligned links at the receiver (bottom).
The plots shown were obtained for both channel models and the 3GPP antenna element, with $N_t=256$, $N_r=64$. Similar plots were obtained for the ISO antenna elements, but were omitted for brevity. Random values of misalignment angles $\Delta_1$ and $\Delta_2$ were generated to obtain the empirical probability distribution functions. 
Furthermore, 
by looking at the support of the distribution in
Figs.~\ref{fig:3gpp-alignfit}--~\ref{fig:misalignedatrx}, it can be
 observed how significant the impact of partial alignment can
be: the probability distribution is indeed spread over significantly
higher values of $G$ for the partially a-aligned link with respect to
the misaligned case, although about 20\,dB smaller than the fully
a-aligned case.

\subsection{System-level validation of the channel approximations \label{subsec:-validation}}
We now validate the above approximations
at the system level. To do so, we
consider the realistic mmwave vehicular network described in Sec.~\ref{subsec:system-trace}.
However, due to the complexity of
fully simulating the 3GPP and the
NYU channel models, we only consider a $0.4 km\times 0.4 km$
area in the center of Luxembourg city, i.e., 8 gNBs and around 120-140 vehicles
served at each 1\,s-time step. 

\begin{figure} 
\centering
\includegraphics[width=0.22\textwidth]{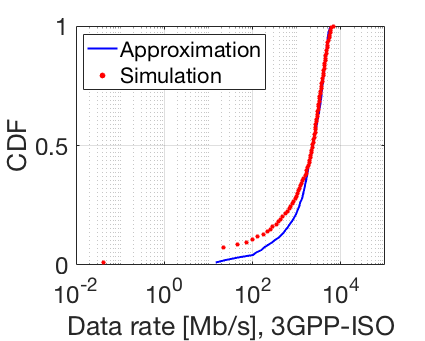}
\includegraphics[width=0.22\textwidth]{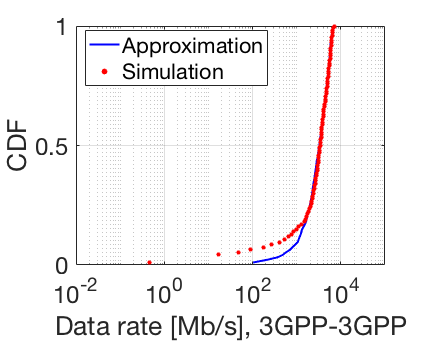}
\includegraphics[width=0.22\textwidth]{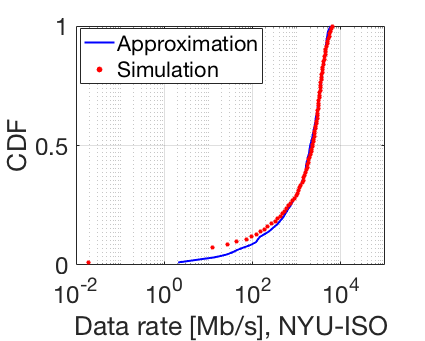}
\includegraphics[width=0.22\textwidth]{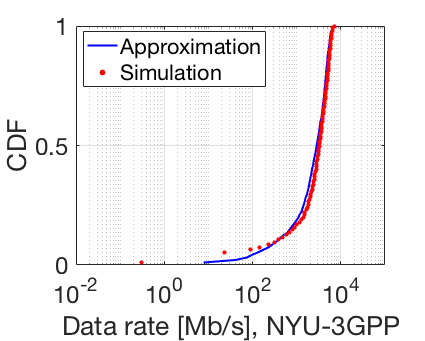}
\caption{CDF of  the  achievable 
  rate obtained via system-level simulations of the Luxembourg
  scenario, for the 3GPP (top) and the NYU (bottom) channel model, and for ISO (left) and 3GPP (right) antennas. The approximation values were obtained applying the models in Secs.~\ref{subsec:fullyaligned3gpp} and \ref{subsec:partiallyaligned}.\label{fig:syslevel-validation-3gpp}}
\end{figure}

Each gNB, equipped with $N_t=256$
antennas, enables a single beam with a static direction chosen
at random. Each vehicle is instead equipped with $N_r=64$ antennas and is
associated to the closest gNB. To quantify the network perfomance, we calculate the signal-to-interference-and-noise ratio
(SINR) for each vehicle $v$ as:
\begin{equation}
\label{eq:sinr-unicas}
{\rm SINR}_v=\frac{P(g^{\star})|\tilde{h}_c(g^{\star},v)|^2}{N_0+\sum_{g\in\Gc\land g\neq g^{\star}}P(g)|\tilde{h}_c(g,v)|^2}
\end{equation}
where $g^{\star}$ is the gNB to which the vehicle is associated, and
$N_0$ is the white noise power. 
The achievable data rate is then: 
\begin{equation}
\label{eq:R-unicas}
  R(b,k,v)= B_w  \log\left(1+{\rm SINR}_v\right)
\end{equation}
where $B_w$ is the channel bandwidth.  The values of data rate we
obtained through Monte Carlo simulations of the channel coefficients are
compared to the values obtained using the approximated channel
gains. More specifically, the approximate gains $G(b,v)$  in~\eqref{eq:gain} are generated for
each gNB-beam-vehicle triple, by drawing random values from the
distributions obtained in the previous subsections, depending on the
level of alignment between their respective beams and LoS
conditions. Fig.~\ref{fig:syslevel-validation-3gpp} shows the
cumulative distribution function (CDF) curves of the 
data rate values obtained under the 3GPP (top) and the NYU (bottom) channel model,
both for the sectored 3GPP and ISO antenna elements. In both
cases, the curves derived under full 3GPP channel simulation fit
well with our proposed approximations, which leverage the distributions
derived in Secs.~\ref{subsec:fullyaligned3gpp} and
\ref{subsec:partiallyaligned}.  Similar results were
  obtained for the SINR, but were omitted  for brevity. These plots show that
the  data rate 
achieved under the approximated channel models, closely matches that 
recorded under the actual 3GPP and NYU  models, with negligible
deviations for the low-SINR vehicles. Importantly, {\em these results allow us
to substantially simplify the mmwave channel model, while maintaining
an adequate level of realism in our scenario}.
 
\section{gNB Selection and Beam Design}
\label{sec:strategies}

We now leverage  the channel gain model described above and  focus
on the main aspects of traffic delivery in mmwave vehicular networks and on the approach we propose to overcome the existing hurdles. In particular, Sec.\,\ref{subsec:goals} 
introduces the beam design problem and our optimization formulation, formally stating its objective and constraints. Then 
\Sec{sub-strategies} presents the heuristic approaches we investigate, 
among which our proposed scheme, TL.

\subsection{Optimization formulation}
\label{subsec:goals}
\label{sec:sub-goals}

Given the system model presented above, here we aim to answer the
following questions: {\em i) how many beams should be active at each gNB, and
of what beamwidth; ii) which directions should they point at on the
azimuth plane}\footnote{Setting the elevation angle is not a goal 
of the optimization, as it depends on the vehicle-gNB distance and 
the antenna tilting angle at the gNB.}{\em; and,
finally, iii) which user}\footnote{It is fair to assume that the beam at the vehicle 
points to the gNB from which the strongest signal is currently
received.} {\em should be scheduled on which beam}. The goal is
to find answers to these questions that maximize the total 
users' data rate. 

In the following section, we present an optimization formulation of the
beam management problem, to be solved at every time step $k$.

For each gNB~$g$, we are given the maximum number~$N(g)$ of beams that 
can be created, their maximum half-power width~$A(g)$ (which is equal to the
sector amplitude in the case of 3GPP antennas), as well as the total
power budget~$P_{tot}(g)$ that has to be split across the gNB beams. 
We also know the angles, $\dv(g,k,v)$, representing the azimuth and the
elevation of vehicle~$v$ as seen from the 
 gNB~$g$ at time~$k$ (see Sec.\,\ref{subsec:channel-model} and Fig.\,\ref{fig:angles}).

To make our  decisions, we have to set many  
variables. Binary variable~$\gamma(b,g,k)$ expresses
whether beam~$b$ is used (i.e., emitted) by gNB~$g$ at time~$k$; such
a dynamic relationship between gNBs and beams allows us to represent
the fact that the same gNB may have a different number of beams at
different times. Continuous variables~$\alpha(b,k)$,
$\delta_1^{(g)}(b,k)$,
and~$P(b,k)$, respectively express the half-power width, directions, and
power of beam~$b$ at time~$k$. Furthermore, for each pair of
beams~$(b_1,b_2)$ with $b_2\neq b_1$, we can use~$b_1$ to generate constructive
interference (CoMP-like) with~$b_2$, or to avoid destructive
interference by keeping it silent (ABS-like), which is modeled through binary
variables~$\mu(b_1,b_2,k)$ and~$\beta(b_1,b_2,k)$, 
respectively. E.g., $\mu(b_1,b_2,k)=1$~means that, at
time~$k$, beam~$b_1$ is used to improve the data rate of~$b_2$ instead of transmitting data on its own. Finally, continuous variables~$\sigma(b,k,v)\in[0,1]$ express the fraction of time that beam~$b$ allocates to serve vehicle~$v$ at time~$k$.

To begin with, we  impose that each beam only belongs to one gNB at
every time interval, and that gNBs do not exceed the maximum number of
beams. I.e., for any $k\in\Kc$,
\begin{equation}
\label{eq:one-gnb}
\sum_{g\in\Gc}\gamma(b,g,k)\leq 1,\,\,\,\forall b\in\Bc; \quad
\sum_{b\in\Bc}\gamma(b,g,k)\leq N(g),\,\,\,\forall g\in\Bc\,.
\end{equation}
In particular, the first inequality descends from the fact that, in our model,
beams in~$\Bc$ are not statically tied to a specific gNB; thus,  the
same element of~$\Bc$ cannot be associated with  multiple gNBs at the same time.

Furthermore, we must ensure that the half-power width of the beams does not exceed the limit, and that beams of the same gNB do not overlap:
\begin{equation}
\label{eq:max-ampl}
\alpha(b,k)\leq A(U(b,k)),\quad\forall b\in\Bc,k\in\Kc,
\end{equation}
\begin{multline}
\label{eq:no-overlap}
|\delta_1^{(g)}(b_1,k)-\delta_1^{(g)}(b_2,k)|\geq \frac{\alpha(b_1,k)+\alpha(b_2,k)}{2},\\
\forall b_1,b_2\in\Bc\colon U(b_1,k)=U(b_2,k),k\in\Kc.
\end{multline}
In both \Eq{max-ampl} and \Eq{no-overlap}, $U(b,k)=g\in\Gc\colon \gamma(b,k,g)=1$ represents the gNB to which beam~$b$ belongs.

\begin{table} 
\caption{Notation used in the formulation of the optimization problem
    \label{tab:notation2}
} 
\scriptsize
\begin{tabularx}{\columnwidth}{|l|l|X|}
\hline
{\bf Symbol} & {\bf Type} & {\bf Meaning} \\
\hline
\hline
$A(g)$ & parameter & Max half-power width of beams of gNB~$g$\\
\hline
$\alpha(b,k)$ & continuous var. & Half-power width of beam~$b$ at $k$\\
\hline
$\sigma(b,k,v)$ & continuous var. & Fraction of time step $k$ that beam~$b$ uses to serve vehicle~$v$\\
\hline
$\mu(b_1,b_2,k)$ & binary var. & Whether beam~$b_1$ supports in a
                                     COMP-like fashion beam~$b_2$ at $k$\\
\hline
$\beta(b_1,b_2,k)$ & binary var. & Whether beam~$b_1$ supports in
                                       an ABS-like fashion beam~$b_2$
                                       at $k$\\

\hline
$\gamma(b,g,k)$ & binary var. & Whether beam~$b$ is used by gNB~$g$ at $k$\\
\hline
$U(b,k)$ & auxiliary var. & gNB using beam~$b$ at $k$\\  
\hline
$V(b,k,v)$ & auxiliary var. & Whether beam~$b$ covers vehicle~$v$
                                  at $k$\\  
\hline
$R(b,k,v)$ & output & Data rate between beam~$b$ and vehicle~$v$ at $k$\\
\hline
$P(b,k)$ & continuous var. & Power of beam~$b$ at $k$\\  
\hline
\end{tabularx}
\end{table}

Concerning transmission power, we mandate that the per-gNB power
budgets are met, and that no power is allocated to unused beams, i.e., 
\begin{equation}
\label{eq:power-budget}
\sum_{b\in\Gc\colon U(b,k)=g}P(b,k)\leq P_{tot}(g),\quad\forall g\in\Gc,k\in\Kc,
\end{equation}
\begin{equation}
\label{eq:no-power-unused}
P(b,k)\leq \sum_{g\in\Gc}\gamma(b,g,k)P_{tot}(g),\quad\forall b\in\Bc,k\in\Kc.
\end{equation} 
Finally, any beam $b$ can be used at any time $k$ in a COMP- or
ABS-like fashion in combination with at another beam:
\begin{multline}
\label{eq:one-comp-abs}
\sum_{b^\prime\in\Bc}\left[\mu(b,b^\prime,k)+\beta(b,b^\prime,k)+\mu(b^\prime,b,k)+\beta(b^\prime,b,k)\right]\leq 1\,.
\end{multline}

Concerning the scheduling variables, we proceed as follows. 
The sum of variables $\sigma(b,k,v)$  cannot exceed one:
\begin{equation}
\label{eq:scheduling}
\sum_{v\in\Vc}\sigma(b,k,v)\leq 1,\quad\forall b\in\Bc,k\in\Kc.
\end{equation}
Additionally, we schedule no time to serve vehicles that are out of the beam's coverage:
\begin{equation}
\label{eq:sched-los}
\sigma(b,k,v)\leq V(b,k,v),\quad\forall b\in\Bc,k\in\Kc,v\in\Vc.
\end{equation}
In \Eq{sched-los},
$V(b,k,v)=\ind{\{|\delta_1^{(g)}(b,k)-d(g,k,v)|\leq\alpha(b,k)/2\}}$
(with $g=U(b,k)$) represents whether beam~$b$ covers vehicle~$v$ at time~$k$.

Our objective can then be stated as maximizing the total data rate received
by vehicles:
\begin{equation}
\label{eq:obj}
\max_{\alpha,\beta,\gamma,\delta_1^{(g)},\mu,\pi,\sigma}\sum_{b\in\Bc}\sum_{k\in\Kc}\sum_{v\in\Vc}\sigma(b,k,v)R(b,k,v) 
\end{equation}
where $R(b,k,v)$ represents the data rate that vehicle~$v$ would obtain from beam~$b$ at time~$k$, if that beam would serve no other vehicle.
Given the availability of the channel coefficients at the gNB, during time
step $k$, i.e., $\Hm_k$, and of the ${\rm SINR}_v$ in
\eqref{eq:sinr-unicas}  experienced by the receiver, the gNB can transmit at the maximum data rate supported by the channel. Then, considering
the  very high waveform spectral occupancy in 5G, 
the achievable data rate through beam
$b$ at time step $k$
is given by \eqref{eq:R-unicas}.
In the case of a CoMP-like
transmission, the useful signal (interference) is 
determined by the number of gNB contributing constructively
(destructively), which are captured by the $V(b,k,v)$ variables.
Note  that, in this case, the total useful signal is given by the
sum of signals coming from different gNBs, which coordinate among
themselves and apply a precoding matrix in order to perform in-phase
transmissions.
It is also important to underline that $R(b,k,v)$ is the estimated
achievable rate, while the actual data rate is determined by the
instantaneous channel conditions and the modulation-coding scheme that
is selected.

\subsection{Beam design heuristics}
\label{sec:sub-strategies}

Directly solving the optimization
problem stated in \Sec{sub-goals} has a very high computational complexity.
Indeed, in both the unicast and brodcast cases, $R$ in the objective
 is a non-linear function of the decision variables, and some of the 
 constraints, e.g., \Eq{no-overlap} and \Eq{sched-los},  involve an
 indicator function and/or the modulo operator, thus making the 
 problem fall in the mixed-integer linear programming (MILP)
 category. 
Such problems are well-known to be NP-hard (the reader is referred to the reduction from the vertex cover problem in~\cite{milp}).

To avoid dealing with such an overwhelming complexity, we adopt a
heuristic approach.

{\bf Clustering-based strategies}. 
 We first cast the beam management 
problem  
as a one-dimensional clustering\footnote{Note
  that {\em clustering} here does not refer to the path clusters
   in Sec.~\ref{subsec:channel-model}. } Indeed, each
beam~$b$ of gNB $g$ can be seen as a cluster of directions, with
center 
$\delta_1^{(g)}(b,k)$ and  maximum width $A(g)$. 
Each of such directions may cover one or more users at a given data
rate, depending on the channel conditions and the antenna gain. Thus,
each direction can be associated with a certain ``weight'' reflecting
the sum rate offered to the covered users. Clustering will lead to
maximizing the total weight corresponding to a set of directions 
(beams). 
We then solve the clustering problem by considering two different
strategies, named {\em Static} and  {\em Dynamic},  as 
detailed next.

In the {\em Static} scheme, the directions and widths of the beams at the gNBs do not
change over time, i.e., $\delta_1^{(g)}(b,k)=\delta_1^{(g)}(b)$ and
$\alpha(b,k)=\alpha(b)$. To 
determine those directions, we formulate a {\em hierarchical
  clustering} problem as follows:
\begin{enumerate}
    \item for each vehicle $v$ and time step $k$, we create one observation (i.e., one data point) corresponding to $v$'s position;
    \item we consider the whole set of observations and compute the
      pairwise {\em angular distances} (i.e.,
      $|\delta_1^{(g)}(b,k)-d(g,k,v)|$ between any two observations;
    \item we feed the resulting distance matrix to the Voor Hees algorithm~\cite{vorhees}, setting the  maximum intra-cluster distance to~$A(g)$;
    \item we consider the~$n$ largest clusters, i.e., the clusters
      including the highest number of observations;
    \item for each such cluster, we set the
      direction~$\delta_1^{(g)}(b)$ as the mean between the minimum and
      maximum angle of the vehicles it includes,
      i.e.,~$\delta_1^{(g)}(b)\gets\frac{1}{2}[\min_{v\in
        \Vc}d_1^{(g)}(v,k)+\max_{v\in\Vc}d_1^{(g)}(v,k)]$.
    \end{enumerate}
    Note that, in step~1, we may create multiple observations with
    the same coordinates, e.g., if two vehicles are observed in the
    same position at different times. This is intentional and allows
    us to properly account for the fact that vehicles are more likely to
   be  found  in some locations of the topology than in others.

The Static strategy is the simplest way to leverage aggregate
traffic statistics; also, it can be performed offline and
requires no reconfiguration of the beams. On the negative side, it
cannot account for the time evolution of vehicular mobility, e.g.,
different traffic patterns at different hours of the day.

The {\em Dynamic} strategy works in the same way as the Static one, with the
important difference that decisions are re-made {\em at every time
step}, i.e., the clustering procedure described above is repeated at
every $k$, accounting only for the positions of the vehicles at that
step. 
Implementing the Dynamic strategy requires real-time knowledge of
vehicular mobility and continuous and  almost-instantaneous beam reconfiguration. Such
aspects reduce its practical relevance to real-world
implementation; nonetheless,  it represents a useful benchmark to
compare against.

{\bf The traffic Lights (TL) strategy}. 
Vehicular mobility is
constrained not only by the road topology, but also by the state of
traffic signals, e.g., traffic lights. Based on such an observation, the TL
strategy leverages the available information on traffic lights states, and
points the beams available at each gNB towards the road segments
where the traffic light  is red. 
Also, all beams will have half-power width equal to $A$.

The TL strategy is more flexible than the Static one, in that beam
directions account for the vehicles' mobility. Also, it
is much more practical than the Dynamic strategy, as beam
reconfigurations are less frequent, and it 
does not require any real-time mobility information. Indeed, it is important to stress
that, unlike the clustering-based Static and Dynamic schemes,
the TL strategy requires {\em no knowledge} whatsoever on
vehicular mobility, and can therefore be applied in situations where
such information is unavailable or unreliable.

\begin{table}
\caption{Toy scenario with two gNBs: quantity of downloaded data, in TByte, under the TL and the optimum strategy
    \label{tab:opti}}
\centering
\begin{tabular}{|l|r|r|}
\hline
Scenario & TL & Optimum \\
\hline\hline
$N=2$, $A=5$ & 14.04 & 14.18 \\
\hline
$N=2$, $A=15$ & 15.20 & 15.26 \\
\hline
$N=4$, $A=5$ & 12.93 & 13.60 \\
\hline
$N=4$, $A=15$ & 14.53 & 14.70 \\
\hline
\end{tabular}
\end{table}

\section{Numerical Results}
\label{sec:results}

In our performance study, we consider the Luxembourg road topology and
mobility trace introduced in Sec.\,\ref{subsec:system-trace}. We then set the carrier frequency at 
 $f_c=76$\,GHz, as typically used in vehicular networks
\cite{comm-radar}, and the available bandwidth at 
$B_w=400$\,MHz \cite{3gpp-nr}. We assume that  all gNBs are equipped with a
$16\times 16$ UPA with up to 4 RF chains, and
the user is equipped with a $8 \times 8$ UPA. 
From the Luxembourg trace \cite{lust}, we utilize:
\begin{itemize}
\item the real-world topology, including building shapes;
\item the real-world location and phases of traffic lights;
\item the realistic traffic and mobility of each vehicles.
\end{itemize}
Furthermore, we adapt the LoS probability used in our model to the
real-world topology we consider, by tailoring the parameters of
the blockage model in \cite{akdeniz-mmwave}.

The channel gains and performance indicators are derived through numerical simulations, based on the above-mentioned real-world data. In particular, we use the 3GPP approximated channel model accounting  
for the Doppler effect, shadowing, and multipath fading, and we set
the large-scale parameters used for modeling as in \cite{3gppchanmodel}.
For all strategies, we consider a common value of maximum half-power
beamwidth  for all gNBs, namely, 
$A=5^{\circ}, 10^{\circ}$ unless otherwise specified, while the
maximum number of
simultaneous beams, $N(g)$, is the same for all the gNBs 
and  varies from 2 to 4. 

We focus on downlink traffic and derive the
  effective data rate by using the 4-bit channel quality indicator (CQI) table in
  \cite{3gpp-tech}, which maps the reported CQI to a particular modulation coding scheme (MCS)
  and spectral efficiency value. For the purposes of this study, the
  SINR to CQI mapping was performed using the spectral-efficiency based
  approach reported in \cite{mezzavilla-cqi}. The values of
data rate depend on the number of gNBs contributing to the useful
received signal (constructive interference), i.e., on
$\mu(b_1,b_2,k)$, as well as on the destructive interference that may
come from other gNBs, i.e., $1-\beta(b_1,b_2,k)$.

\begin{figure}
\centering
\includegraphics[width=.3\textwidth]{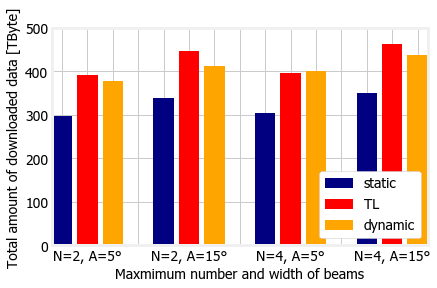}
\caption{Total amount of downloaded data for different strategies and
  beam configurations, (i.e., maximum number
  of beams, beam half-power width).\label{fig:totalthr}}
\end{figure}

The first aspect we are interested in is the performance of the
proposed TL strategy against the optimum. Due to the problem
complexity, the latter is found via brute-force, and, to make the
brute-force approach terminate in a reasonable amount of time, 
we limit the comparison to two gNBs, namely, the two that in the
topology in \Fig{scenario} are closest to each other.

\Tab{opti} presents the total amount of data that the vehicular users
download under the TL and optimum strategies, and accounting for the
actual channel conditions the users experience. We observe that TL is
remarkably close to the optimum; furthermore, the difference is more
significant in four-beam scenarios than in two-beam ones. This is due
to the fact that, in some four-beam configuration, under TL  the same
vehicle may be served by multiple beams, which does not happen under the optimal strategy.

 We also investigate the performance of TL  in the full
Luxemburg City scenario and compare it against the other strategies discussed in
\Sec{sub-strategies}. These results, shown 
in \Fig{totalthr}, have been obtained by considering again the actual channel
conditions over time. As one might expect, more and/or wider
beams consistently result in better performance. 
The comparison between different strategies is instead quite
surprising. For all but one set-up, TL outperforms all other
schemes, including Dynamic which uses real-time information (the
reasons for this behavior are detailed next).  
As for Static, it performs fairly well in comparison to the
alternatives, proving to be not only an interesting baseline but also
a viable option 
when more sophisticated approaches are infeasible, e.g., when gNBs cannot be co-located with traffic lights.

\begin{figure}
\centering
\includegraphics[width=.23\textwidth]{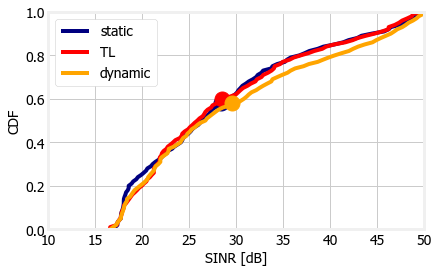}
\includegraphics[width=.23\textwidth]{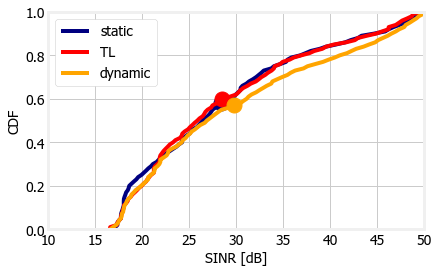}
\caption{SINR experienced by served vehicles under different strategies, when there are: two beams (left) and four
  beams (right) with $A=5^{\circ}$. The average values are marked by a dot on each curve.
    \label{fig:sinr}
}
\centering
\includegraphics[width=.23\textwidth]{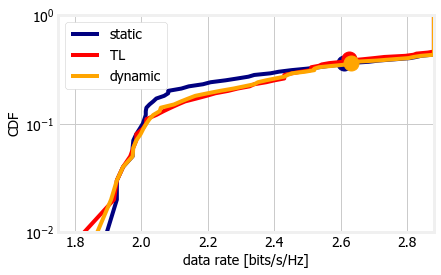}
\includegraphics[width=.23\textwidth]{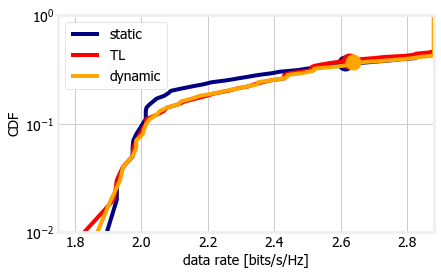}

\caption{Actual rate experienced by served vehicles under different strategies, when there are: two beams (left) and four
  beams (right) with $A=5^{\circ}$. The average values are marked by a dot on each curve.
    \label{fig:rate}
} 
\end{figure}

To understand where the difference in performance we
observed in \Fig{totalthr} comes from, we look at 
SINR and data rate achieved by the
different strategies, summarized in \Fig{sinr} and \Fig{rate}, 
respectively.   We can observe that Dynamic is able to achieve a better
SINR than its alternatives; the average SINR value (marked by the dot in the curve) is around 2~dB higher for Dynamic than the other two strategies. This is thanks to the fact that beams can be
steered exactly towards vehicles instead of towards road
segments. Unsurprisingly, Static has the lowest SINR, owing to the
fact that the direction of beams cannot change over time. 
There is another interesting effect we can observe when comparing
\Fig{sinr} and \Fig{rate}. The Static strategy results in a slightly
lower SINR for the vehicles experiencing worse conditions (the ones representing the worst 30\% in the CDF curve), as we can see from
the left part of the curves in \Fig{sinr}.  Such a difference
corresponds, in \Fig{rate}, to a lower rate for those vehicles. The
Dynamic strategy, on the other hand, provides the vehicles enjoying
favourable propagation conditions  with  a substantially better SINR
than its alternatives, as we can observe form the right part of the
curves in \Fig{sinr}. 
However, the difference in rate in \Fig{rate} is much more limited; the average rate values, marked by the dots on the curves, almost coincide for the three different strategies. The reason is that SINR values are often so high that vehicles achieve the maximum possible rate, as shown by the jumps in \Fig{rate}: further increasing the SINR, as Dynamic does, brings little additional benefit. 
Comparing the left and right plots in \Fig{sinr} and \Fig{rate}, we can conclude that increasing the number of beams from 2 to 4, the system performance does not improve significantly.

\begin{figure}
\centering
\includegraphics[width=.23\textwidth]{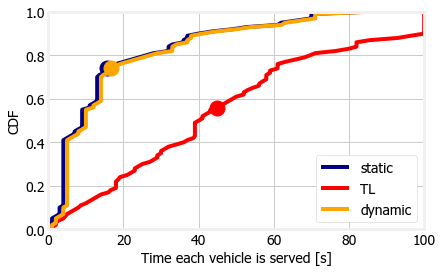}
\includegraphics[width=.23\textwidth]{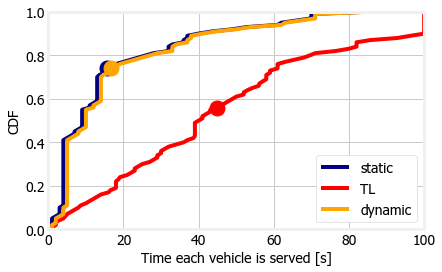}
\caption{Time for which each vehicle is served under different
  strategies, for  two beams (left) and four beams (right)
  with $A=5^{\circ}$. The average values are marked by a dot on each curve.
    \label{fig:vehtime}
} 
\centering
\includegraphics[width=.23\textwidth]{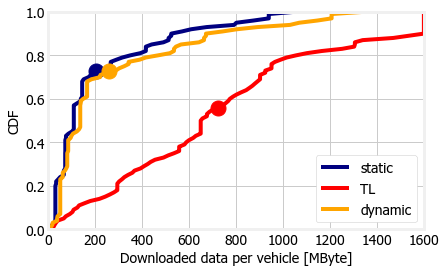}
\includegraphics[width=.23\textwidth]{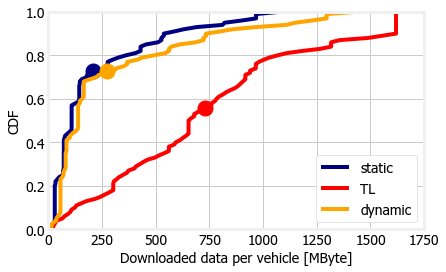}
\caption{Data downloaded by each served vehicle under
  different strategies, when there are: two beams (left) and four
  beams (right) with $A=5^{\circ}$. The average values are marked by a dot on each curve.
    \label{fig:vehdata}
} 
\end{figure}

The differences in SINR and actual rate shown in \Fig{sinr} and
\Fig{rate} do not fully explain, however, the performance difference displayed
by \Fig{totalthr}. 
In Figs.\,\ref{fig:vehtime} and \ref{fig:vehdata}, we therefore study for how long each
vehicle is served under each strategy and the amount of
downloaded data per vehicle. 
Recall that, in both figures, only served vehicles are considered. 
The difference between TL and its counterparts is now very clear and
very significant. Under all set-ups, TL serves vehicles for a
substantially longer time than the other strategies (\Fig{vehtime})
and, 
accordingly, it provides them with much more data. 
By comparing \Fig{vehtime} to \Fig{vehdata}, we see
that, although vehicles are served for virtually the same time by the
Static and Dynamic strategies (\Fig{vehtime}), 
the latter results in a higher quantity of transferred
data. In particular, there is an average increase of
  approximately $60$~Mbyte per vehicle;

this is a consequence of the difference in data rate we observed in \Fig{rate}.

The fact that vehicles are served for a much longer time under the TL
strategy is consistent with  the basic idea that TL should serve static vehicles waiting at a red light.

In summary, the TL strategy focuses on the vehicles that can profit the most from mmwave gNBs deployed at traffic lights, and provides them with as much data as possible for as long as possible. This may sound unfair; however, it is worth recalling that mmwave is not supposed to be  the be-all and end-all of vehicular networks (or indeed, any kind of network). Mmwave is one of several technologies to be combined together in order to provide pervasive, high-capacity network coverage to mobile users, and using it to serve static users waiting at traffic lights is, in our scenario, the best way to make the most out of it.

\section{Conclusions}
\label{sec:conclusion}
 Mmwave is a promising technology to enhance the
capacity of vehicular networks. However, the performance of mmwave
networks depends on the number, the alignment, and the width of beams
between gNBs and vehicles, and these require knowledge of the
vehicles' mobility. Instead of relying on real-time mobility
information,  we  proposed to rely on traffic signals, e.g.,
traffic lights, which influence the mobility itself. In particular,
we first  developed low-complexity approximate models for the mmwave
channel gain and showed their accuracy in realistic scenarios, against
more complex, existing models. Then we presented an optimization
formulation for an effective beam design, and we proposed a
low-complexity, heuristic solution, which proved to perform very close
to the optimum.

Our performance evaluation, based on our innovative mmwave communication models and  real-world topology and mobility information, has provided relevant insights. Leveraging traffic light-state information for beam design results in a network performance that exceeds that of baseline approaches (namely, static beam alignment) and is comparable to that of approaches using real-time mobility information.

\bibliography{mmwave}
\bibliographystyle{IEEEtran}
\end{document}